\documentclass[conference,10pt]{IEEEtran}
\usepackage{amssymb,amsmath,theorem}
\usepackage{array}
\usepackage{graphicx}
\usepackage{epsfig}
\usepackage{amsfonts}
\usepackage{bm}
\usepackage{arydshln}
\usepackage{subfig}
\usepackage{algorithm}
\usepackage{algorithmic}
\usepackage{stfloats}

\newcolumntype{C}[1]{>{\centering}p{#1}}
\usepackage{rotating,setspace,latexsym,amsmath,epsf,amssymb,bm,color}
\usepackage{cite}
\usepackage{booktabs}

\IEEEoverridecommandlockouts

\ifCLASSINFOpdf
\else
\fi

\begin{document}
%
\title{3D Cooperative User Tracking for Distributed Integrated Sensing and Communication }

%

\author{\IEEEauthorblockN{Yingjie~Xu\IEEEauthorrefmark{1},
Xuesong~Cai\IEEEauthorrefmark{2},
        Michiel~Sandra\IEEEauthorrefmark{1}, 
        Sara~Willhammar\IEEEauthorrefmark{1}, and~Fredrik~Tufvesson\IEEEauthorrefmark{1}
        }
        
\IEEEauthorblockA{\IEEEauthorrefmark{1}
Department of Electrical and Information Technology, Lund University, 22100 Lund, Sweden.\\
\IEEEauthorrefmark{2}School of Electronics, Peking University, Beijing, 100871, P. R. China.\\
E-mail: \{yingjie.xu, michiel.sandra, sara.willhammar, fredrik.tufvesson\}@eit.lth.se, xuesong.cai@pku.edu.cn.}
}

\maketitle

\begin{abstract}
As integrated sensing and communication (ISAC) becomes an integral part of 6G networks, distributed ISAC (DISAC) is expected to enhance both sensing and communication performance through its decentralized architecture. This paper presents a complete framework to address the challenge of cooperative user tracking in DISAC systems. By incorporating a global probability hypothesis density (PHD) filter and a field-of-view-aware access point (AP) management strategy, the framework enables accurate user tracking using radio signals while optimizing AP scheduling. In addition, a real-world distributed MIMO channel measurement campaign is performed to evaluate the effectiveness of the framework. The results demonstrate that a centimeter-level root mean-square trajectory error can be achieved. Furthermore, the results show that it is not necessary to keep APs active at all times to maintain high tracking accuracy, indicating the need for robust and efficient AP management. These findings provide valuable insights into practical deployments and the further development of cooperative user tracking techniques in DISAC systems.       
\end{abstract}
\begin{IEEEkeywords}
Distributed integrated sensing and communication (DISAC), user tracking, PHD filter, AP management, distributed MIMO channel measurement.  
\end{IEEEkeywords}

%
\IEEEpeerreviewmaketitle

\section{Introduction}
As a key enabling technology for upcoming 6G wireless networks, integrated sensing and communication (ISAC) extends communication systems to include environmental sensing~\cite{H. Guo2024}. Meanwhile, distributed multiple-input multiple-output (MIMO), also known as cell-free MIMO~\cite{H. Q. Ngo 2017}, is emerging as a key candidate for next-generation MIMO architectures. The spatial diversity enabled by the distributed architecture offers significant improvements in both communication and sensing capabilities. As a result, integrating distributed MIMO with ISAC into a unified system, referred to as distributed ISAC (DISAC)~\cite{E. C. Strinati2024,U. Demirhan2023 }, is a promising direction to further explore the full potential of ISAC.

One important sensing aspect of DISAC systems is user equipment (UE) tracking using radio signals, which can be categorized as a multi-target tracking issue. In this context, random finite set (RFS) based tracking methods~\cite{W. Wu2023}, which integrate RFS with a Bayesian filtering framework, have been widely studied. A representative approach is the probability hypothesis density (PHD) filter~\cite{K. Granstrom2012_1}. It can effectively address key challenges during tracking, including uncertain data associations, a time-varying number of targets, missed detections, and false alarms. Based on the PHD filter, various multi-target tracking challenges have been extensively explored in the literature~\cite{S Jovanoska2012,W. Yi2020,K. Granstrom2012_2}. However, most existing studies focus primarily on the tracking with sensor networks, which cannot be directly applied to ISAC systems. In a mobile communication context, a multiple-model PHD filter is designed in~\cite{H. Kim2020_1}, performing UE tracking using millimeter-wave signals. To further account for multipath effects in communication channels, an extended Kalman PHD filter (EK-PHD) and a cubature Kalman PHD (CK-PHD) filter are reported in~\cite{O. Kaltiokallio2024} and~\cite{H. Kim2020_2}, respectively. However, the approaches~\cite{H. Kim2020_1, O. Kaltiokallio2024, H. Kim2020_2} are limited to co-located MIMO setups, with the lack of support for widely distributed MIMO deployments. 

Compared to co-located MIMO setups, distributed MIMO architectures introduce new challenges for UE tracking. Specifically, due to the limited communication range and computational resources, it is neither efficient nor practical for all APs to continuously communicate with the UEs~\cite{U. Demirhan2023}. Therefore, reliable AP management is necessary to determine AP activities at any given time. A common criterion relies on each AP's field of view (FoV)~\cite{J. Kim2019,Y. Ge2024}, that is, an AP is active only when the UE is within its FoV. An FoV-based handover method was proposed in~\cite{Y. Ge2024} for DISAC systems. However, it considers only one single AP at a time, neglecting the potential effectiveness of cooperative communication and sensing among multiple APs. In addition, the method is evaluated solely in simulated environments, without experimental verification of tracking performance in real-world deployments.

To the best of our knowledge, a comprehensive study of UE tracking in DISAC systems, including algorithm design, AP management strategies, and practical performance evaluation, has not been fully explored in the literature. To fill this gap, we provide a complete framework for 3D cooperative UE tracking in DISAC systems, and validate it through practical experiments. The main contributions and novelties of this work are as follows.     

\begin{enumerate}
        \item A framework for UE tracking is designed based on a PHD tracking filter. The framework includes PHD prediction, FoV-aware AP management, measurement transformation, and PHD updating procedures. A key contribution is its ability to enable accurate 3D UE tracking along with dynamic AP management.
        \item 
        We conduct a real-world channel measurement campaign in which distributed APs are deployed throughout a room, and $128\times780$ multi-link channels are measured along a 12-m UE route.
        \item The effectiveness of the proposed framework is validated using the measured channels. Compared to scenarios without AP management and those using other tracking filters, our approach demonstrates significantly lower resource consumption and improved tracking accuracy, respectively.         

\end{enumerate}
The presented framework, along with its experimental evaluation, serves as a valuable reference for implementation and advancement in the area of cooperative UE positioning/tracking in DISAC systems.

\section{System Model}
\subsection{Signal Model}
\begin{figure}[tb]
	\centering
		\begin{minipage}[tb]{0.3\textwidth}
			\centering
			\includegraphics[width=1\textwidth]{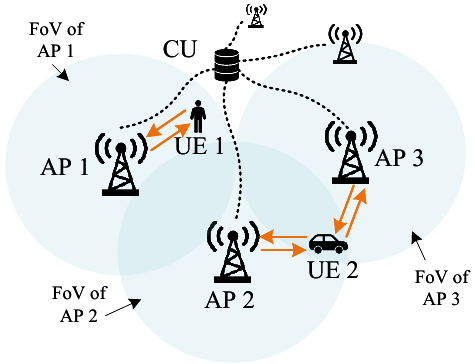}
		\end{minipage}
	\caption{Multi-user distributed MIMO system.}
	\label{fig:DISAC system}
\end{figure}

As shown in Fig.~\ref{fig:DISAC system}, consider a distributed MIMO system where $K$ distributed APs serve $M$ single-antenna users. Each AP is assumed to have a limited service range, defined as its FoV, and the associated communication link is established only when a UE is within the AP's FoV. Assume that all APs are synchronized with the UE, i.e. no clock bias issues. Furthermore, assume that the APs can communicate synchronously with the UEs. At the $k$-th AP, given the transmit signal $\mathbf{s}_{k,m}$ from the $m$-th UE, the received signal is expressed as\footnote{Without loss of generality, uplink communication is considered here.} 
\begin{equation}
\begin{split}
    \mathbf{y}_{k,m}(t)=&{\textstyle \sum_{l=1}^{L_{k,m}}}\mathbf{c}^{r}\left(\mathbf{\Omega}_{l,k,m}^{r}\right)\mathbf{A}_{l,k,m} e^{-j2\pi v_{l,k,m}t}\\
    & \cdot \mathbf{s}_{k,m}\left(t-\tau_{l,k,m}\right)+\mathbf{n}_{k,m},
    \label{eq:signal model}
\end{split}
\end{equation}
where $L_{k,m}$, $\mathbf{c}^{r}$, and $\mathbf{n}_{k,m}$ are denoted as the number of multipath components (MPCs) in the channels, the antenna response on the AP side, and the additive Gaussian noise, respectively. Each MPC is characterized with a delay $\tau_{l,k,m}$, an angle of arrival $\mathbf{\Omega}_{l,k,m}^{r}$, a polarization matrix $\mathbf{A}_{l,k,m}$, and a Doppler frequency $v_{l,k,m}$. The angle vector $\mathbf{\Omega}_{l,k,m}^{r}$ is defined by both the azimuth angle $\phi_{l,k,m}^{r}$ and the elevation angle $\theta_{l,k,m}^{r}$, and is expressed as $\mathbf{\Omega}_{l,k,m}^{r}=\begin{bmatrix} \cos \phi_{l,k,m}^{r} \sin \theta_{l,k,m}^{r}   & \sin \phi_{l,k,m}^{r} \sin \theta_{l,k,m}^{r}  &  \cos\theta_{l,k,m}^{r} \end{bmatrix} ^{\text{T}}$.
\subsection{State Model}
The state of the $k$-th AP contains its 3D location $\mathbf{p}^{\text{AP}}_k$ and 3D orientation $\mathbf{r}_k^{\text{AP}}$, which are considered prior knowledge. Assuming that the UEs move over time, the state $\mathbf{x}_{m}^t$ of the $m$-th UE at time step $t$ is characterized by its 3D position $\mathbf{p}^{\text{UE}}_{m}(t)$ and its 3D velocity $\mathbf{\upsilon}_{m}^t$. The transition density is expressed as 
\begin{equation}
f_{t+1|t}\left(\mathbf{x}_{m}^{t+1}|\mathbf{x}_{m}^t\right)=\mathcal{N}\left(\mathbf{x}_{m}^{t+1};\zeta \left(\mathbf{x}_{m}^t\right),\mathbf{Q}^{t+1}\right),
\label{eq:transition model}
\end{equation}
where $\zeta(\cdot)$ and $\mathbf{Q}$ denote the transition model (e.g. a constant-velocity model) and the process noise covariance, respectively.  
\begin{algorithm}[tb!]
	\renewcommand{\algorithmicrequire}{\textbf{Input:}}
	\renewcommand{\algorithmicensure}{\textbf{Output:}}
\caption{Measurement transformation using sigma points}
\label{algorithm2}
\begin{algorithmic}[1]
\REQUIRE(For clarity, the indices of the MPC $l$, AP $k$, UE $m$, and time instant $t$ are omitted here.) Range-bearing measurements $\tilde{\mathbf{z}}$ and the covariance matrix $\tilde{\mathbf{w}}$. 
\ENSURE Position-related measurements $\mathbf{z}$ with the measurement covariance matrix $\mathbf{w}$.
\STATE Factorize $\tilde{\mathbf{w}}=\Sigma \Sigma^{\text{T}}$.
\STATE Given the dimension $d_{\mathbf{z}}$ of $\tilde{\mathbf{z}}$, define $\mathbf{e}_{\mathbf{z},c}$ as the $c$-th column vector of the matrix $[\mathbf{I}_{d_\mathbf{z},d_\mathbf{z}},-\mathbf{I}_{d_\mathbf{z},d_\mathbf{z}}]\in \mathbb{R}^{d_\mathbf{z}\times 2d_\mathbf{z}}$, $c=1,...,2d_{\mathbf{z}}$. 
\STATE Calculate the $c$-th cubature point for $\tilde{\mathbf{z}}$ as 
$\mathbf{z}_c \doteq[\tau_{c},\theta_{c},\phi_{c}]^{\text{T}}=\Sigma \delta_{\mathbf{z},c}+\tilde{\mathbf{z}},$
where $\delta_{\mathbf{z},c}=\sqrt{d_{\mathbf{z}}}\mathbf{e}_{\mathbf{z},c}$.
\STATE Evaluate the propagated cubature point using  
$\mathbf{u}_{c}=\mathbf{r}^{\text{AP}} \cdot \begin{bmatrix} \cos\phi_c\sin\theta_c&  \sin\phi_c\sin\theta_c & \cos\theta_c \end{bmatrix}^{\text{T}}.$
\STATE The associated position-related measurement is given by $\mathbf{z}=\frac{1}{2d_\mathbf{z}}\sum_{c=1}^{2d_{\mathbf{z}}}\mathbf{u}_c$
with the measurement covariance matrix $\mathbf{w}=\frac{1}{2d_\mathbf{z}}\sum_{c=1}^{2d_{\mathbf{z}}}\mathbf{u}_c\mathbf{u}_c^{\text{T}}-\mathbf{z}\mathbf{z}^{\text{T}}$.

\end{algorithmic}			
\end{algorithm}
\subsection{Measurement Model}
To solve a UE tracking problem, we consider a two-stage processing as suggested in~\cite{W. Wu2023}. First, after receiving the signals, channel estimation is performed at each AP. Second, the estimated MPC delay and AoA are obtained and treated as \emph{range-bearing measurements} for the subsequent tracking. Assuming that the measurements of different APs are independent and homogeneous, they are transformed into  \emph{position-related measurements} $\mathbf{z}_{l,k,m}^t$ in the state space. This is achieved using the inverse sigma points~\cite{A. F. Garcia-Fernandez2015}, which is common practice to linearize non-linear measurements and is detailed in Algorithm~\ref{algorithm2}.
Next, the resulting measurement model for the UE tracking is expressed as
\begin{equation}
\mathbf{z}_{l,k,m}^t = \mathbf{p}^{\text{UE}}_{m}(t) + \mathbf{w}^{t}_{l,k,m},
\label{eq:measurement model}
\end{equation}
where $\mathbf{w}^{t}_{l,k,m}$ denotes the measurement noise with a covariance of~$\mathbf{W}^{t}_{l,k,m}$. The resulting likelihood is modeled as
\begin{equation}
h\left(\mathbf{z}_{l,k,m}^t|\mathbf{x}_{m}^t\right)=\mathcal{N}\left(\mathbf{z}_{l,k,m}^t;\xi\left(\mathbf{x}_{m}^t\right),\mathbf{W}^{t}_{l,k,m}\right),
\end{equation}
with $\xi\left(\mathbf{x}_{m}^t\right)=\mathbf{p}^{\text{UE}}_{m}(t)$.
Note that there are also non-line of sight (NLoS) paths in the received measurements, i.e., MPCs which may have experienced reflection, scattering, or diffraction during propagation. In this work, such paths are treated as clutter\footnote{Note that there are available methods that exploit NLoS paths for UE tracking, as reported in~\cite{H. Kim2020_1,O. Kaltiokallio2024,H. Kim2020_2, Y. Ge2024}, which is not considered in our work.}. Furthermore, it is unclear which measurements originate directly from the UEs or from clutter, making it difficult to perform an accurate data association. To address this, we model all the measurements received by the $k$-th AP at time $t$ as an RFS, i.e., $\mathbf{Z}_{k}^t=\left\{ \mathbf{z}_{1,k}^t,...,\mathbf{z}_{L,k}^t\right\}$, $L=\sum_{m}^{M}\hat{L}_{k,m}$, and the UE states at time $t$ as another RFS $\mathbf{X}^t=\left\{ \mathbf{x}_{1}^t,...,\mathbf{x}_{M}^t\right\}$. Note that $\hat{L}_{k,m}$ may not be equal to the true MPC order $L_{k,m}$ in~(\ref{eq:signal model}) due to false alarms or missed detections. Since LoS paths may also be undetected in some time steps, the detection probability $p_{\text{D}}\in[0,1]$ is introduced to characterize the possibility that a measurement is directly received from a UE. Based on this model, we employ a PHD tracking filter, where RFS $\mathbf{Z}_{k}^t$ is used to estimate the UE state RFS $\mathbf{X}^t$, as will be introduced in the following section.

\section{3D Cooperative UE Tracking Framework}
\begin{figure}[tb]
	\centering
		\begin{minipage}[tb]{0.46\textwidth}
			\centering
			\includegraphics[width=1\textwidth]{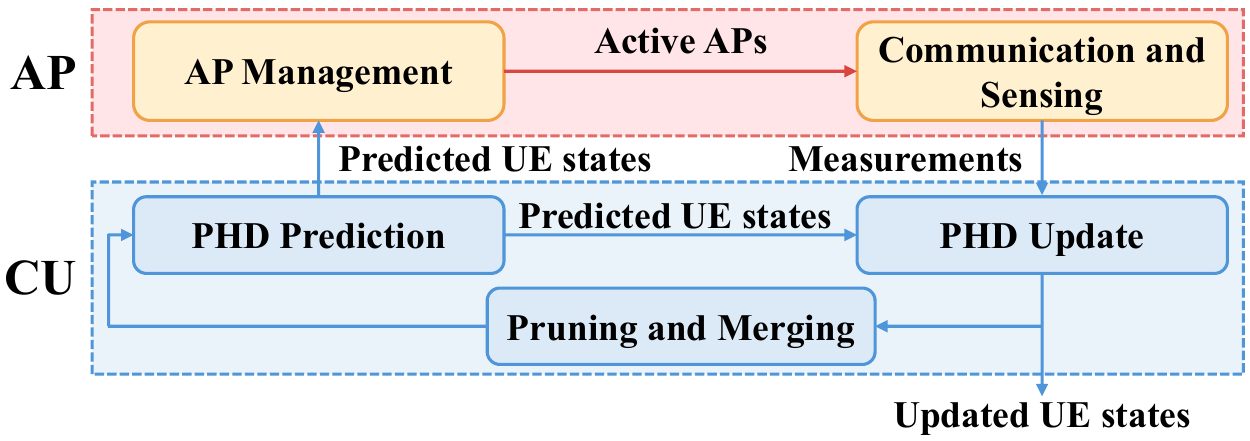}
		\end{minipage}
	\caption{The workflow of the complete 3D cooperative UE tracking framework.}
	\label{fig:workflow}
\end{figure}

In this section, our 3D cooperative UE tracking framework is introduced. All APs are assumed to be connected to a central control unit (CU), which collects and integrates the measurements $\mathbf{Z}_{k}^t$ from the APs. At each time step, different APs may be active for communication with UEs and sharing their measurements with the CU, according to a predefined AP management strategy. Using the shared measurements, a global PHD filter is performed in the CU to enable smooth tracking of the UEs. A workflow of the complete framework is shown in Fig.~\ref{fig:workflow}. In the following subsections, we first present the implementation of the global PHD filter, followed by the details of the presented AP management strategy.      

\subsection{Global PHD Filter}
Based on the Bayesian filtering framework, a PHD filter aims to recursively estimate the first-order statistical moment of the RFS posterior density $p(\mathbf{X}^t|\mathbf{Z}^t_k)$, referred to as the PHD $v_t(\mathbf{x})$. The local maxima of $v_t(\mathbf{x})$ contain information about the UE-state RFS $\mathbf{X}^t$ and can be used to estimate the individual UE states, i.e., the elements of $\mathbf{X}^t$. Specifically, the PHD can propagate over time through the following prediction and updating steps: 
\begin{equation}
\begin{split}
v_{t|t-1}(\mathbf{x}) = \int f_{t|t-1}(\mathbf{x}|\mathbf{x}^{t-1}) v_{t-1}(\mathbf{x}^{t-1}) \text{d} \mathbf{x},
\label{eq:PHD rescure1}
\end{split}
\end{equation}
\begin{equation}
\begin{split}
    v_t(\mathbf{x}) =& (1 - p_D) v_{t|t-1}(\mathbf{x}) \\
    & + \sum_{\mathbf{z} \in \mathbf{Z}^t_k} \frac{p_D h_t(\mathbf{z}|\xi(\mathbf{x})) v_{t|t-1}(\mathbf{x})}{\lambda_c + p_D \int h_t(\mathbf{z}| \xi(\mathbf{x})) v_{t|t-1}(\mathbf{x}) \text{d}\mathbf{x}},
    \label{eq:PHD rescure2}
\end{split}
\end{equation}
where $f(\cdot)$, $h(\cdot)$, and $\lambda_c$ denote the transition density, likelihood function and clutter intensity, respectively. To derive a closed-form solution to (\ref{eq:PHD rescure1}) and (\ref{eq:PHD rescure2}), we model the PHD as a Gaussian mixture (GM)\cite{K. Granstrom2012_1}, namely,
\begin{equation}
    v_t(\mathbf{x}) = \sum_{i=1}^{J(t)} w_i^t \mathcal{N}\left(\mathbf{x}; \mathbf{m}_i^t, \mathbf{P}_i^t\right),
\end{equation}
with $J(t)$, $w^t$, $\mathbf{m}^t$, and $\mathbf{P}^t$ denoting the number of Gaussian components, the component weight, the component mean, and the component covariance at time $t$, respectively. Accordingly, the prediction step~(\ref{eq:PHD rescure1}) can be rewritten as
\begin{equation}
    v_{t|t-1}(\mathbf{x}) = \sum_{i=1}^{J_{t|t-1}} w_i^{t|t-1} \mathcal{N}\left(\mathbf{x}; \mathbf{m}_i^{t|t-1}, \mathbf{P}_i^{t|t-1}\right),
    \label{eq:GM-PHD1}
\end{equation}
with $w_i^{t|t-1}=w_i^{t-1}$, $\mathbf{m}_i^{t|t-1} = \zeta(\mathbf{m}_i^{t-1})$, and
$\mathbf{P}_i^{t|t-1} = \zeta\left(\mathbf{m}_i^{t-1}\right)\mathbf{P}_i^{t-1} \left[\zeta\left(\mathbf{m}_i^{t-1}\right)\right]^{\text{T}}+\mathbf{Q}^{t-1}.$
The update step~(\ref{eq:PHD rescure2}) can be rewritten as
\begin{equation}
\begin{split}
    v_t(\mathbf{x}) &= (1 - p_D) v_{t|t-1}(\mathbf{x})+\sum_{\mathbf{z} \in \mathcal{Z}^t}\sum_{i=1}^{J_{t|t-1}}w^{t|t}_i\left(\mathbf{x};\mathbf{m}_{i}^{t|t},\mathbf{P}_{i}^{t|t}\right),
    \label{eq:GM-PHD2}
\end{split}
\end{equation}
where
\begin{equation}
\begin{split}
     w^{t|t}_i = \frac{p_Dw^{t|t-1}_i\mathcal{N}\left(\mathbf{z};\xi\left(\mathbf{m}^{t|t-1}_{i}\right),\mathbf{S}_{i}^{t}\right)}{\lambda_c + p_D \sum_{i}^{J_{t|t-1}}w^{t|t-1}_i\mathcal{N}\left(\mathbf{z};\xi\left(\mathbf{m}^{t|t-1}_{i}\right),\mathbf{S}_{i}^{t}\right)},
\end{split}
\end{equation}
\begin{equation}
\mathbf{m}_i^{t|t}=\mathbf{m}_i^{t|t-1}+\mathbf{K}_i^{t}\left(\mathbf{z}^{t}-\xi\left(\mathbf{m}_i^{t|t-1}\right)\right),
\end{equation}
\begin{equation}
\mathbf{P}_i^{t|t}=\mathbf{P}_i^{t|t-1}-\mathbf{K}_i^t\mathbf{S}_i^t\left[\mathbf{K}_i^t\right]^\text{T},
\end{equation}
\begin{equation}
\mathbf{K}_i^{t}=\mathbf{P}_i^{t|t-1}\xi\left(\mathbf{m}_i^{t|t-1}\right)^\text{T}\left[\mathbf{S}_i^t\right]^{-1},
\end{equation}
\begin{equation}
\mathbf{S}_i^{t}=\xi\left(\mathbf{m}_i^{t|t-1}\right)\mathbf{P}_i^{t|t-1}h
\left(\mathbf{m}_i^{t|t-1}\right)^\text{T}+\mathbf{W}^t.
\label{eq:16}
\end{equation}
After the update step, the estimated UE state RFS $\hat{\mathbf{X}}^t$ at time $t$ is determined by taking the mean $\mathbf{m}_i^{t|t}$ of the Gaussian components with the $M$-largest weights $w_i^{t|t}$, i.e.,
\begin{equation}
\hat{\mathbf{X}}^t=\left\{\mathbf{m}_i^{t|t}:i\in \mathbf{I}=\arg_{\{ i_1,...,i_M\}}w_{i_1}^{t|t}\geq w_{i_2}^{t|t}\geq...\geq w_{i_{J_{t}}}^{t|t}\right\}.
\label{eq:UE state estimation}
\end{equation}
Note that the number of GM components tends to grow exponentially over time. To reduce computational complexity, pruning and merging techniques are applied after each PHD update. Specifically, GM components with small weights are discarded, and those with similar parameters are merged and represented as a single GM component~\cite{B.-N. Vo2006}.  

It is worth noting that a key feature of the presented filter is the multi-source measurement transformation before each PHD update. That is, range-bearing measurements from different APs are first transformed into position-domain measurements in the same state space. Then we adopt a clustering method in~\cite{T. Li2017} to group those multi-source position-domain measurements corresponding to the same objects and generate a `proxy' measurement used in PHD update. This differs from existing filters~\cite{H. Kim2020_1,O. Kaltiokallio2024,H. Kim2020_2}, in which range-bearing measurements are used directly in the nonlinear measurement model, leading to the need to approximate them through a nonlinear filter that generally exhibits high computational complexity.

\subsection{FoV-Aware AP Management}
A strategy to determine which APs are active at each time step is essential to maintain continuous UE tracking while minimizing system overhead. A basic selection criterion can be based on the FoVs of each AP~\cite{J. Kim2019,Y. Ge2024}. In an DISAC system, the FoV can be defined as a region in which an AP can effectively provide sensing and communication services, considering both physical distance and angular direction. APs are only active when the UEs are within their own FoVs. Given this criterion, we introduce FoV-aware AP management during the tracking process.  

Let $\Gamma_t$ denote the set of AP indices whose FoVs contain at least one UE at time $t$. The measurements $\mathcal{Z}^t=\{\mathbf{Z}_k^t:k\in \Gamma_t \}$ collected by these APs in $\Gamma_t$ are uploaded to the CU. Let $\zeta_{\Gamma_{t}}=\{v_{t+1|t}(\mathbf{x}),w^{t+1|t},\mathbf{m}^{t+1|t},\mathbf{P}^{t+1|t}\}_{\Gamma_t}$ represent the predicted tracking information from the APs in $k\in \Gamma_t$. The predicted UE states for the time $t+1$ are determined from $\hat{\mathbf{X}}^{t+1|t}=\{\mathbf{m}_i^{t+1|t}:i\in \mathbf{I}=\arg_{\{ i_1,...,i_M\}}w_{i_1}^{t+1|t}\geq w_{i_2}^{t+1|t}\geq...\geq w_{i_{J_{t}}}^{t+1|t}\}$. To determine APs that are expected to communicate with UEs at time $t+1$, we define a FoV-aware selection criterion as 
\begin{equation}
\Gamma_{t+1}=\{k: \hat{\mathbf{p}}^{\text{UE}}_m(t+1|t) \text{ is in }A_{k}, m=1,...,M \},
\label{eq:active APs}
\end{equation} 
where $A_k$ denotes the FoV of the $k$-th AP, which is typically determined by the beam pattern of the AP and its communication signal-to-noise ratio~(SNR). For example, when using directional antennas, the FoV of an AP can be modeled as a limited circular sector, and (\ref{eq:active APs}) can be reformulated as 
\begin{equation}
\begin{split}
\Gamma_{t+1}=&\{k:\left \|  \hat{\mathbf{p}}^{\text{UE}}_m(t+1|t)-\mathbf{p}_k^{\text{AP}}    \right \|_{\text{F}}\leq d_{th}\} \\
&\cap  \{k: \text{angle}\{ \hat{\mathbf{p}}^{\text{UE}}_m(t+1|t)-\mathbf{p}_k^{\text{AP}},\mathbf{r}^{\text{AP}}_k\}\leq \theta_{th} \}  .
\label{eq:active APs2}
\end{split}
\end{equation}
Here, $d_{th}$ and $\theta_{th}$ characterize the effective radiation range of the AP, $\text{angle}\{\mathbf{a},\mathbf{b}\}$ represents the angle between the vectors $\mathbf{a}$ and $\mathbf{b}$. For omnidirectional antennas, the angular condition in~(\ref{eq:active APs2}) can be omitted. Once $\Gamma_{t+1}$ is determined, the work of sharing measurements with the CU will be handed over from the APs $k \in \Gamma_{t}$ to the APs $k \in \Gamma_{t+1}$. The subsequent update of the UE state (\ref{eq:PHD rescure2}) is then performed with the measurements $\mathcal{Z}^{t+1}=\{\mathbf{Z}_k^{t+1}:k\in \Gamma_{t+1} \}$.  

In summary, based on the predicted UE states at every time step, $\Gamma$ is formed by selecting one or multiple APs whose FoVs cover the predicted UE positions. Then, the work of sensing and sharing measurements with the CU is handed over to the newly selected APs. The complete PHD tracking algorithm with FoV-aware AP management is described in Algorithm~\ref{algorithm1}. 
\begin{algorithm}[tb!]
	\renewcommand{\algorithmicrequire}{\textbf{Input:}}
	\renewcommand{\algorithmicensure}{\textbf{Output:}}
\caption{PHD tracking algorithm with FoV-aware AP Management}
\label{algorithm1}
\begin{algorithmic}[1]
\REQUIRE AP states $\{\mathbf{p}_k^{\text{AP}},\mathbf{r}_k^{\text{AP}} \}$, measurements $\{\mathbf{Z}_k^t \}$, the number of APs $k=1,...,K$, and time step $t=1,...,T$. 
\ENSURE UE states $\hat{\mathbf{X}}^t, t=1,...T$. 
\STATE Initialize UE states $\hat{\mathbf{X}}^{(0)}$ and $\Gamma_{0}$.
\WHILE{$t\leq T$}
\STATE Calculate $\zeta_{\Gamma_{t-1}}$ and $\hat{\mathbf{X}}^{t|t-1}$ through the prediction step~(\ref{eq:GM-PHD1}).
\STATE Initialize $\Gamma_{t}\leftarrow \emptyset $, $\mathcal{Z}^{t}\leftarrow \emptyset$.
\FOR{$m\in M$, $k\in K$}
\IF{$\hat{\mathbf{p}}_m^{\text{UE}}(t|t-1)$ is in $A_k$}
\STATE AP handover: $\Gamma_{t}\leftarrow\Gamma_{t}\cup k$.
\STATE Measurements sharing: $\mathcal{Z}^{t}\leftarrow\mathcal{Z}^{t}\cup \mathbf{Z}_k^{t}$.
\ENDIF
\ENDFOR
\STATE Perform updating step~(\ref{eq:GM-PHD2}) and calculate $\hat{\mathbf{X}}^{t|t}$ through~(\ref{eq:UE state estimation}).
\STATE Pruning and merging.
\STATE $t\leftarrow t+1$
\ENDWHILE
\end{algorithmic}			
\end{algorithm}
\section{Experimental Analysis}
\subsection{Indoor Distributed MIMO Channel Measurements}
\begin{figure}[t]
	\centering
	\subfloat[]
	{
		\begin{minipage}[tb]{0.21\textwidth}
			\centering
			\includegraphics[width=1\textwidth]{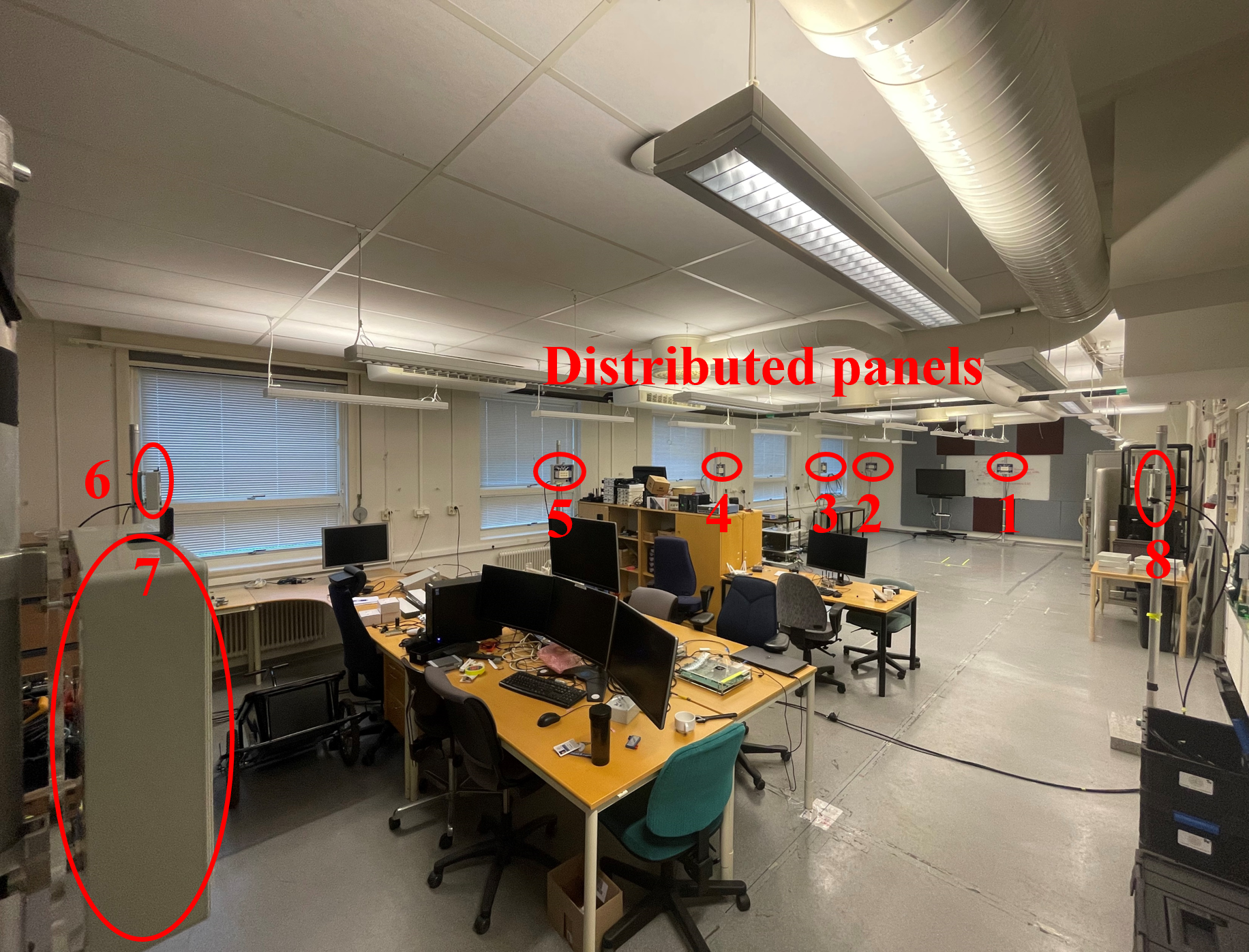}
		\end{minipage}
	}
    	\subfloat[]
	{
		\begin{minipage}[tb]{0.25\textwidth}
			\centering
			\includegraphics[width=1\textwidth]{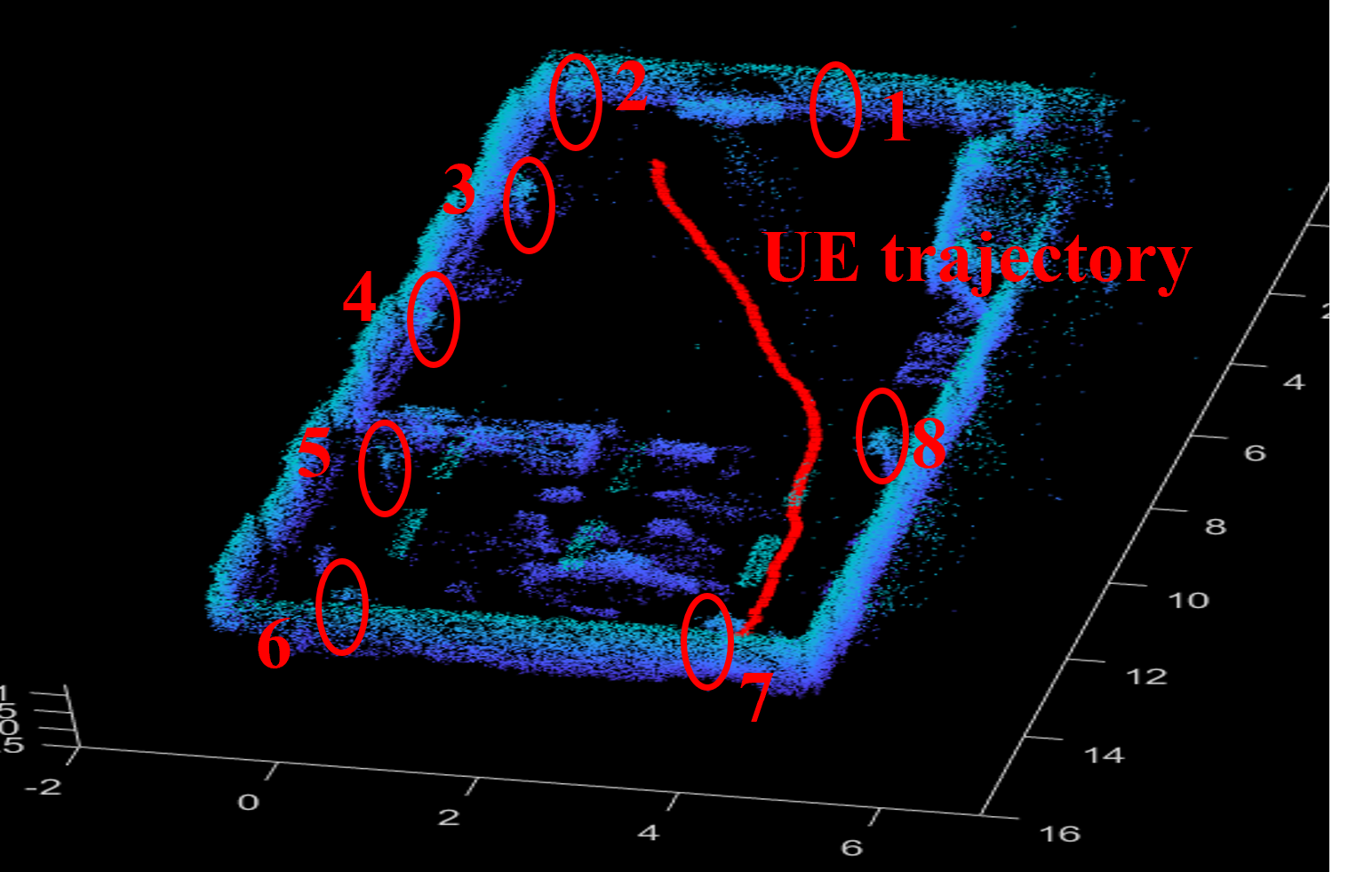}
		\end{minipage}
	}
	\caption{(a) The measurement environment and (b) ground truth of the mobile UE.}
	\label{fig:Measurement_environments}
\end{figure}

To validate the performance of the proposed framework, a distributed MIMO channel measurement campaign was carried out, as illustrated in Fig.~\ref{fig:Measurement_environments}a. A wideband switch-based channel sounder~\cite{M. Sandra2024} was deployed for the measurements. On the BS side, eight 16-element uniform planar arrays (UPAs), referred to as `panels', were employed. On the UE side, a single omnidirectional antenna was used. Before measurements were made, a back-to-back calibration was performed to eliminate the responses of the sounder hardware, connectors, and cables. 

The measurements were performed with a carrier frequency of 5.6~GHz and a bandwidth of 400~MHz. The panels were placed in fixed positions throughout the room, as shown in Fig.~\ref{fig:Measurement_environments}(a). The UE antenna was mounted on a mobile robot, which moved from one side of the room to the other, as depicted in Fig.~\ref{fig:Measurement_environments}b. A global coordination system was established with its origin set at the UE start position. The ground truth of the UE trajectory was accurately recorded by a Lidar sensor. In total, the uplink channels between eight panels and 780~discrete UE positions (corresponding to 780 time steps) were measured, resulting in the collection of $780\times128$ channel snapshots. 

The MPC parameters, including delay, AoA, Doppler frequency, and polarization matrix, were extracted from the measured channel impulse responses using the SAGE algorithm~\cite{X. Yin 2003}. The estimated number of MPCs in the algorithm is set to 10, which was found to offer a good trade-off between accurately capturing the LoS paths and minimizing computational complexity. It should be noted that, due to practical limitations, the proposed framework was evaluated in a single-UE distributed MIMO communication scenario. A more comprehensive performance analysis involving multi-user scenarios is left for future work.

\subsection{Results and Analysis}
\begin{figure}[tb]
	\centering
		\begin{minipage}[tb]{0.43\textwidth}
			\centering
			\includegraphics[width=1\textwidth]{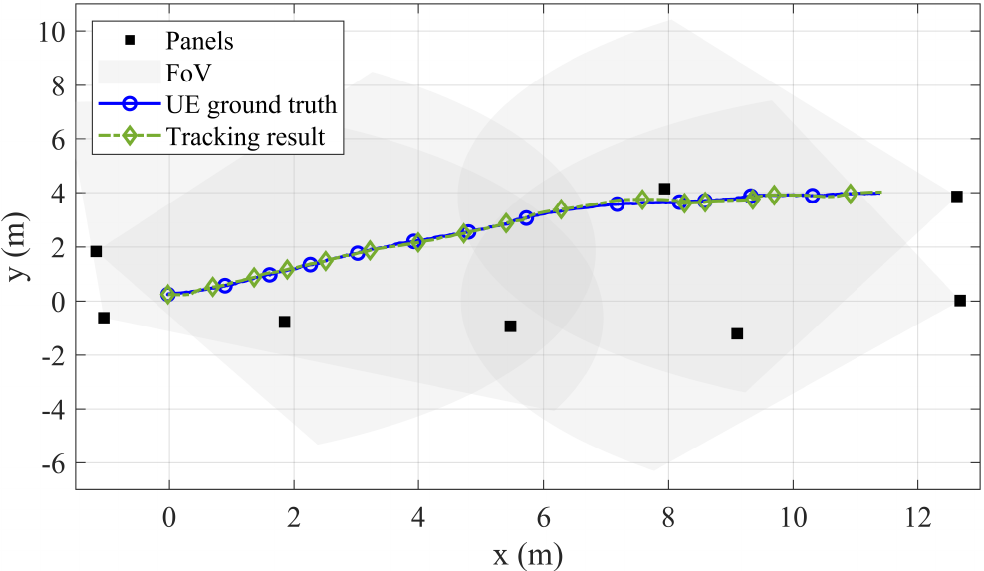}
		\end{minipage}
	\caption{Field of views of the panel 1,~2,~6, and~7, ground truth of the UE trajectory, and the tracking result with the proposed framework.}
	\label{fig:tracked route}
\end{figure}

Given the measured channels, UE tracking is performed using the proposed framework. Within the framework, a random walk transition model is applied in~(\ref{eq:transition model}). The process noise covariance is set as $\mathbf{Q}=\text{diag}[0.1^2 \text{ m}^2,0.1^2 \text{ m}^2,0.1^2 \text{ m}^2]$. 
The maximum number of Gaussian components is set to 500, with the pruning threshold and merging threshold set to $10^{-4}$ and 4, respectively. To define the FoVs of the panels, $d_{th}$ and $\theta_{th}$ are set at 8.5~m and $60^\circ$ respectively, which effectively characterize the main-lobe direction range of the panels and the region where reliable SNR is achievable~\cite{M. Sandra2024}. As examples, the resulting FoVs of the panel~1, 2, 6, and 7 are depicted in Fig.~\ref{fig:tracked route}. 

\begin{figure}[tb]
	\centering
		\begin{minipage}[tb]{0.45\textwidth}
			\centering
			\includegraphics[width=1\textwidth]{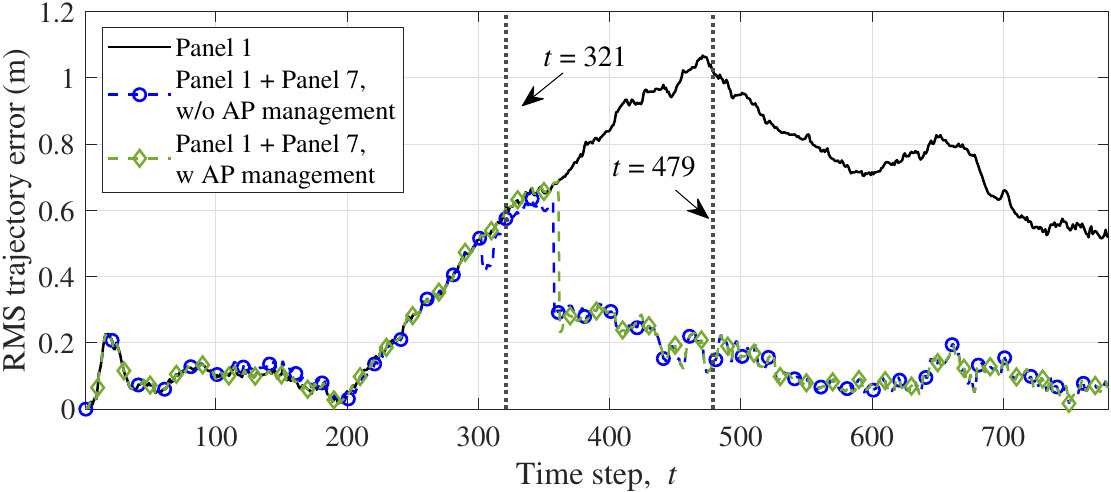}
		\end{minipage}
	\caption{RMS trajectory errors of the UE tracking based on panel~1 only, panel~1 and panel~7 with the introduced AP management strategy, and panel~1 and panel~7 without AP management.}
	\label{fig:rmse p1p7}
\end{figure}

The root mean squared (RMS) trajectory error~\cite{Y. Ge2024} is studied to evaluate the performance of the UE tracking. To better analyze the effectiveness of the introduced AP management strategy, we first select panel~1 and panel~7 for the tracking of the UE. The RMS trajectory error over time is presented in Fig.~\ref{fig:rmse p1p7}. For comparison, tracking is also performed using only measurements from panel~1, and the corresponding RMS trajectory error is included in the figure. Before time step~321, the UE remains within the FoV of panel~1 only, resulting in the same tracking performance in both scenarios. After this, the UE enters the FoV of panel~7. As panel~7 becomes active and begins to contribute to the measurements, the RMS trajectory error decreases compared to the case using only panel~1. After time step~479, the UE moves out of the FoV of panel~1, and the task of sensing and communication is fully handed over to panel~7. By inspecting the tracking performance of panel~1, we can conclude that such a handover is necessary. After the UE exits the FoV of panel~1, continued reliance on its measurements leads to reduced tracking accuracy due to increased communication distance, and thus to degraded SNR. In contrast, the tracking based on panel~7 demonstrates improved accuracy during this period. Additionally, we investigate the scenario in which both panels remain active throughout the tracking process, i.e., without any AP management. The results show that the use of two fully active panels provides only marginal improvement in accuracy. These findings demonstrate the importance of appropriate AP management, which enables continuous and accurate UE tracking while minimizing system overhead. 

\begin{figure}[t!]
	\centering
		\begin{minipage}[tb]{0.45\textwidth}
			\centering
			\includegraphics[width=1\textwidth]{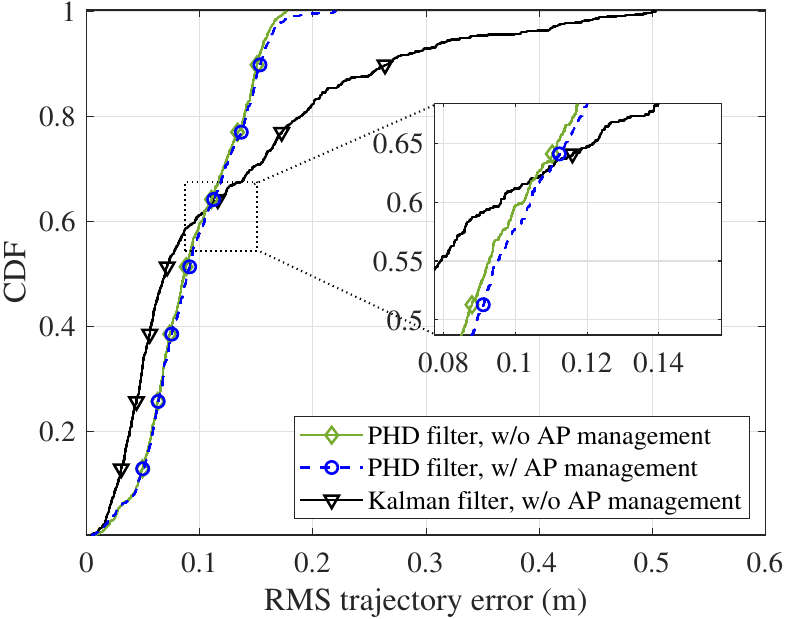}
		\end{minipage}
	\caption{CDFs of RMS trajectory errors.}
	\label{fig:rmse cdf}
\end{figure}

\begin{figure}[t!]
	\centering
		\begin{minipage}[tb]{0.45\textwidth}
			\centering
			\includegraphics[width=1\textwidth]{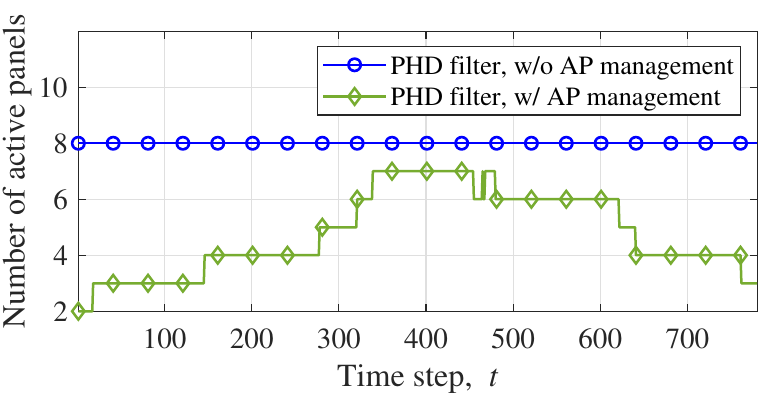}
		\end{minipage}
	\caption{Number of active panels over time step}
	\label{fig:num of active panels}
\end{figure}

Furthermore, leveraging the proposed framework, UE tracking is performed with all available panels. The tracked result is visualized in Fig.~\ref{fig:tracked route}. In addition, the CDF of the RMS trajectory error is shown in Fig.~\ref{fig:rmse cdf}. The results demonstrate good tracking performance with an average error of 0.09~m, the tracking is also performed without AP management, where all panels remain active and continuously contribute to the measurements. The corresponding CDF of the RMS trajectory error is also shown in Fig.~\ref{fig:rmse cdf}. The results indicate that the proposed algorithm achieves nearly the same tracking performance as the full-time active approach. It is important to note that the FoV-aware AP management strategy significantly reduces the number of panels that operate simultaneously. Specifically, the number of active panels over time is shown in Fig.~\ref{fig:num of active panels}. The average number of active panels is 4.78, indicating a more efficient resource allocation and improved energy efficiency.

Finally, to further evaluate the effectiveness of the PHD filter in the proposed framework, a standard Kalman filter (KF)-based tracking method is implemented for comparison. In the KF approach, the data association is performed using the nearest-neighbor principle, representing a hard decision-based association strategy. The parameters required in the KF align with those employed in the proposed method. The resulting CDF of the RMS trajectory error is shown in Fig.~\ref{fig:rmse cdf}. It can be seen that the KF-based method exhibits limited tracking performance. This degradation is primarily due to incorrect data association, particularly when the LoS path is not detected. Such situations are common in practice due to potential blockages in the environment, instability in channel measurements, or limited accuracy of channel parameter estimation.
In contrast, the PHD filter used in the proposed method treats all measurements collectively as a random finite set, thereby avoiding the need for explicit data association. As a result, it demonstrates greater robustness in real-world scenarios.
\section{Conclusions}
In this paper, cooperative UE tracking in DISAC systems has been investigated. Using a global PHD filter and FoV-aware AP management, a novel framework has been proposed to achieve accurate UE tracking and dynamic AP scheduling. A distributed MIMO channel measurement campaign has been conducted to evaluate the practical performance of the proposed framework. The results have shown that the proposed framework achieves centimeter-level RMS trajectory error. Meanwhile, FoV-aware management maintains high tracking accuracy while significantly reducing the number of active APs at any given time, thus decreasing the overall overhead of the system. The proposed framework and its experimental evaluation offer a valuable reference for implementation and advancement in cooperative positioning and tracking in DISAC systems.  
\section*{Acknowledgment}
This work was supported by the Swedish strategic research area ELLIIT, by NextG2Com funded by the VINNOVA program for Advanced Digitalization with grant number 2023-00541, and also partially funded by 6GTandem, supported by the Smart Networks and Services Joint Undertaking (SNS JU) under the European Union's Horizon Europe research and innovation program under Grant Agreement No 101096302.
\ifCLASSOPTIONcaptionsoff
  \newpage
\fi

\end{document}